\documentstyle[proceedings,psfig]{crckapb} 

\newcommand{\mdot}{$\dot{M}$}
\newcommand{\msolyr}{M$_{\odot}$yr$^{-1}$}
\newcommand{\lsol}{L$_{\odot}$}

\newcommand{\more}{\raisebox{-1.1mm}{$\stackrel{>}{\sim}$}}

\begin{opening}
\title{AGB STARS IN THE MAGELLANIC CLOUDS }
\vspace{-0.99cm}

\author{M.~A.~T.~GROENEWEGEN}
\institute{Max-Planck-Institut f\"ur Astrophysik, \\
Karl-Schwarzschild-stra{\ss}e 1, D-85748 Garching, Germany}

\vspace{-0.99cm}
\end{opening}

\runningtitle{AGB STARS IN THE MAGELLANIC CLOUDS}

\begin{document}



\vspace{-0.3cm}
\section{Introduction}

The Magellanic Clouds (MCs) are an ideal testing ground for theories
of the late stages of stellar evolution. They are nearby and yet far
enough that to first order the depth of the MCs may be neglected, and
all stars can be considered to be at the same distance (approximately
50 and 63 kpc for the LMC and SMC, respectively).

One of the primary observables is the luminosity function of
oxygen-rich and carbon-rich AGB stars. Until the late-eighties AGB
stars were searched for in the optical, either spectroscopically
(using grisms or direct spectroscopy of red stars) or by identifying
variable stars in the appropriate period range. Infrared observations
were usually done as follow-up. When the IRAS Point and Faint source
catalogues became available people started 
identifying AGB stars from those catalogues. These searches find
a different population of AGB stars that are more luminous and more
redder and that have been missed by the previous optical searches.

To illustrate this, I have used a dust radiative transfer model to
calculate the expected magnitudes of a short period, low luminosity
and a long period, high luminosity carbon Mira at the distance of the
LMC (see Table~1). The estimates of the luminosities and mass loss
rates are from the period-luminosity relation of Groenewegen \&
Whitelock (1996) and the period-mass loss rate relation of
(Groenewegen et al. 1997).

It is clear that low and intermediate luminous carbon stars (but
similarly for O-rich AGB stars) are bright in the optical but are far
below the detection limit of IRAS, which was approximately 200 mJy at
12 $\mu$m. Stars more luminous than approximately 10~000 \lsol, or less
luminous stars with a very high mass loss rate could have been
detected by IRAS. In addition one should not forget that most AGB
stars are large-amplitude LPV's (long period variables). The
full-amplitude of Miras are \more 2.5 mag in $V$, \more 0.9 mag in $I$
and \more 0.4 mag in $K$ and at 12$\mu$m. IRAS could therefore also have
detected less luminous stars that happened to be near maximum light at
the time of the IRAS observations.

It is clear that both the optical surveys and IRAS must be
biased. From Table~1 it appear that the near-infrared (NIR) region may be
the most unbiased way to search for AGB stars. In fact, there are two
projects underway that will do so, namely the NIR sky-surveys
DENIS and 2MASS.

In addition ISO provides the opportunity the make detailed
observations of individual objects with unprecedented sensitivity and
at wavelengths not accessible from the ground.

In Sect.~2 the surveys that have been conducted 
to identify AGB stars are described. In Sect.~ 3 and 4 first results from
DENIS and ISO are presented. I conclude in Sect.~5.

\begin{table}[htb]
\vspace{-0.5cm}
\begin{center}
\caption{Expected fluxes for carbon Miras in the LMC}
\begin{tabular}{llllllllllll} \hline
Period & Lum. & \mdot & $V$ & $I$ & $J$ & $H$ & $K$ & $[12]$ & $[25]$ \\ 
 (days)& (\lsol) & (\msolyr) & (mag) &  &  &  &  & (mJy) & (mJy) \\
\hline
320    & 3000 & 1.7 $\times$ 10$^{-8}$& 17.2 & 14.4 & 12.5 & 11.7 &
11.0 & 6.0 & 1.6 \\
680    & 10000 & 8.4 $\times$ 10$^{-6}$& 26.8 & 20.9 & 16.1 & 13.9 &
11.7 & 177 & 76 \\
\hline
\end{tabular}
\end{center}
\vspace{-1.0cm}
\end{table}

\section{Surveys}

To construct the luminosity function of AGB stars one must first
identify them, and then perform some follow-up observations to be able
to determine the apparent bolometric magnitude. In the section the
most well-known and recent surveys, and references to follow-up work
are described. I distinguish between LMC, SMC, the inter cloud region
and clusters.

\vspace{-0.2cm}
\subsection{LMC}

\begin{itemize}

\item Westerlund et al. (1978) discovered 302 C-stars over an 62.5
sq. degree area and give coordinates, charts and photographic $I$.
Follow-up observations were performed by Richer et al. (1979) who
present photoelectric $RI$ for 112 of them, and Cohen et
al. (1981) who present $JHK$ photometry for 25.

\item Blanco et al. (1980) discovered 186 C-stars and 102 M5+ stars in
three fields of 0.12 sq. degree each (the `bar west', `optical
center' and `radio center' fields) and present photographic $RI$
photometry, charts and coordinates. Follow-up work was done by Cohen
et al. (1981) who present $JHK$ photometry for 53 and $m_{\rm bol}$
for all carbon stars in the three fields based on bolometric
corrections, and by Richer (1981) who give $VRI$ and $m_{\rm bol}$
for 71 carbon stars in the `bar west' field.

\item Blanco \& McCarthy (1983) discovered 1045 C and 480 M6+ stars in
52 fields of 0.12 sq. degree, which include the Blanco et
al. fields. A total number of 11~000 carbon stars in the LMC is
estimated. Follow-up is presented by Blanco \& McCarthy (1991) who
give charts and coordinates for 849 C-stars, and by Costa \& Frogel
(1996) who present $RI$ photometry for 888 and $JHK$ observations for
204 C-stars, and estimate $m_{\rm bol}$ for all of them.

\item Frogel \& Blanco (1990) present an extended survey for M-giants
in the `bar west' field. They present $JHK$ and $m_{\rm bol}$ for a
sample of 128 M-giants.

\end{itemize}

\noindent
Searches for AGB stars have also been done using the variable star
character of AGB stars. 
The largest survey so far is that by Hughes (1989) who found 471
Miras and 572 SRs over 53 sq. degrees in the LMC. They give mean $I$ magnitudes
and amplitudes, charts, periods and light curves. In a follow-up paper
Hughes \& Wood (1990) present $JHK$ photometry for 267 Miras and 117
SRs, and classify 121 O- and 87 C-stars from optical spectra. Other
recent work includes that of Reid et al. (1995) who find 302 periodic
variables, at least 190 of which are Miras, for which they present
charts and $JHK$ photometry. Sebo \& Wood (1995) identify 19 LPV's near
the cluster NGC 1850.

\vspace{-0.2cm}
\subsection{SMC}
\vspace{-0.2cm}

\begin{itemize}

\item Blanco et al. (1980) found 134 C- and 5 M5+ stars in 2 fields of
0.12 sq. degree each, and give $RI$ photometry, charts and
coordinates. Cohen et al. (1981) present $JHK$ photometry for 20 of them.

\item Blanco \& McCarthy (1983) presented 789 C- and 57 M6+ stars in 37
fields of 0.12 sq. degree each, including the 2 Blanco et al. fields. 
A total number of 2900 carbon stars in the SMC is estimated.

\item Westerlund et al. (1986) discovered 449 C-stars in 2 fields of
0.78 sq. degree each, 405 of which are new discoveries.  They derive
$m_{\rm bol}$.

\item Reid \& Mould (1990) presented a photographic $VI$ survey of 0.8
sq. degree and identify AGB stars candidates from $V-I$
color. Spectroscopic observations confirm  18 C-, 18 M- and 43 K-giants.

\item Rebeirot et al. (1993) presented 1707 C-stars in 13 fields of
0.78 sq. degree each (including the 2 Westerlund et al. fields) and
give charts and coordinates.  Westerlund et al. (1995) present $JHK$ data
for 50, and medium-dispersion spectra for 39 of them.

\item Morgan \& Hatzidimitriou (1995) surveyed 220 sq. degrees of the
outer parts of the SMC. They find 1634 C-stars of which 449 are also
in Rebeirot et al. (1993).

\end{itemize}

\noindent
In Fig.~1 the luminosity function (LF) of carbon stars in the LMC and SMC
is compared. The apparent $m_{\rm bol}$ data for the LMC is taken from 
Costa \& Frogel (1996) and I assumed a distance modulus of 18.5. For
the SMC, I took the 1636 stars from Rebeirot et al. (1993) with
good magnitudes and colors, used the bolometric correction relation from
Westerlund et al. (1986), and assumed a distance modulus of 19.0. 

The two LFs are significantly different. The LF of SMC carbon stars is
much broader although this may be partly due to the bolometric
correction calculation which introduces some spread. The difference in
the peaks of the LFs appears real however. The mean luminosity of
carbon stars in the LMC is about 7000 \lsol, that of their SMC
counterparts about 4300 \lsol. A qualitative explanation may be that
due to the lower metallicity oxygen-rich AGB stars need fewer
dredge-up events to become carbon stars and hence do so at lower
luminosities. Another explanation may be that the population of SMC
stars contains relatively more low mass stars that are intrinsically
fainter. The challenge of AGB population synthesis models (like
Groenewegen \& de Jong 1993, 1994, Groenewegen et al. 1995) is to 
explain this difference quantitatively.

Another interesting feature is the relatively large number of very 
faint carbon stars, with luminosities below the tip of the RGB at
$M_{\rm bol} \approx -3.5$. This was already noted by Westerlund et al. (1992).

\begin{figure}
\vspace{-0.1cm}
\centerline{\psfig{figure=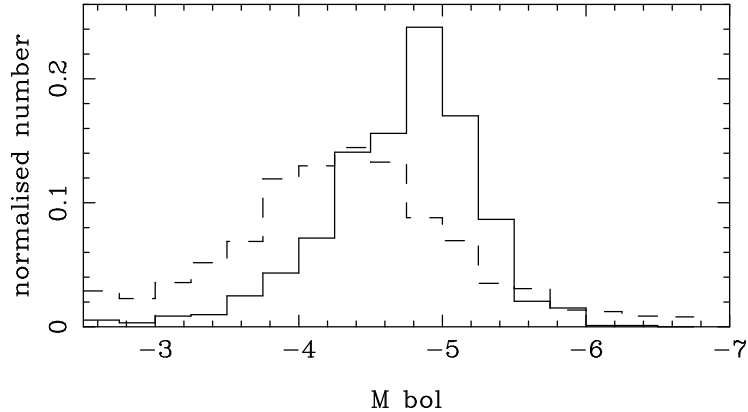,width=10.0cm}}
\vspace{-0.3cm}
\caption[]{Luminosity function of carbon stars in the LMC (solid line;
data from Costa \& Frogel 1996) and SMC (dashed line; derived by me
from data in Rebeirot et al. 1993). First bin contains all stars
fainter than $M_{\rm bol} = -2.5$.
}
\vspace{-0.3cm}
\end{figure}

\vspace{-0.3cm}
\subsection{Inter Magellanic Cloud region}
\vspace{-0.1cm}



Demers et al. (1993) find 57 red stars ($B-V>1.765$). From
spectra they find that 33 are C- and 12 M-stars. Feast \& Whitelock
(1994) present $JHK$ photometry for all of them.

In a recent paper, Kunkel et al. (1997) discover 392 carbon stars in
the outer parts of the SMC and inter cloud region, and give radial
velocities, coordinates and charts.


\vspace{-0.2cm}
\subsection{Magellanic Cloud clusters}

Many of the work on this subject in the early-1980's is by the late 
M. Aaronson, J. Mould and J. Frogel. Recent work is by Frogel, Mould
\& Blanco (1980), Westerlund et al. (1991), Ferraro et al. (1995),
Tanab\'e et al. (1997).

\vspace{-0.2cm}
\subsection{IRAS triggered research}

In the late-eighties a whole new field of MCs AGB research was born
when people started to look for AGB stars in the direction of the MCs
using the IRAS data products. The list below summarises this work.

\begin{itemize}

\item Whitelock et al. (1989) discuss five objects in the direction of the SMC.

\item Reid et al. (1990) and Reid (1991) combined IRAS with optical
data and performed $JHK$ photometry to end up with the identification of
ten ``cocoon'' stars in the direction of the LMC.

\item Wood et al. (1992) provided detailed observations of 3 objects in
the SMC and 16 in the LMC. Six show OH-maser emission which confirms
they are oxygen-rich and allows their terminal wind velocity to be
determined. The sources are monitored in the infra-red from which
pulsation periods in range 930-1390 days are found for nine of
them. This are periods much longer than for optical LPVs.

\item Groenewegen et al. (1995) presented the first ground-based 
8-13 $\mu$m spectra of two AGB stars in the MCs, one in each
cloud. These happened to have about the same pulsation period and by
comparing with a galactic OH/IR source of roughly the same period, it was
found that the ratio of the dust optical depth was Galaxy:LMC:SMC =
15:10:1. This suggested a ratio of mass loss rates of 5:4:1 and hence
the first, albeit rough, quantitative estimate of the dependence of
mass loss rate on metallicity.

\item Zijlstra et al. (1996), Loup et al. (1997a), van Loon et
al. (1997a,b) is a series of paper by essentially the same group of people.

Loup et al. classify 91 IRAS sources in the SMC and 635 in the LMC: 59
are optically known AGB or RSG (red supergiants); 36 are 
confirmed obscured AGB/RSG stars based on NIR photometry of
Zijlstra et al. and the other papers listed above; 23 are PNe; 209 are
candidate AGB/RSG stars; 91 are ruled out as AGB stars, 164 are
foreground objects and 154 could be any type of objects.

van Loon et al. present additional and follow-up observations in the
optical, near- and mid-infrared. Work in progress consists of
determining the pulsation periods, and obtaining 3 $\mu$m spectra to
classify the objects.

\item Groenewegen \& Blommaert (1997) identify 29 IRAS AGB candidates
in the direction of the SMC. NIR photometry has been obtained
and some optical spectra and photometry as well.

\end{itemize}
\vspace{-0.2cm}

\noindent
Based on literature data and new preliminary pulsation periods for
IRAS detected AGB stars available early 1996, we have made model fits
to the spectral energy distributions (SEDs) to obtain the mean luminosity and
plotted them in a period vs. luminosity diagram to compare them to the
$P-L$-relations that had been derived for short period Miras. The
result is in Fig.~2 (taken from Groenewegen et al. 1996). The conclusion 
is that the IRAS stars appear to be on extensions of these relations.

\begin{figure*}
\centerline{\psfig{figure=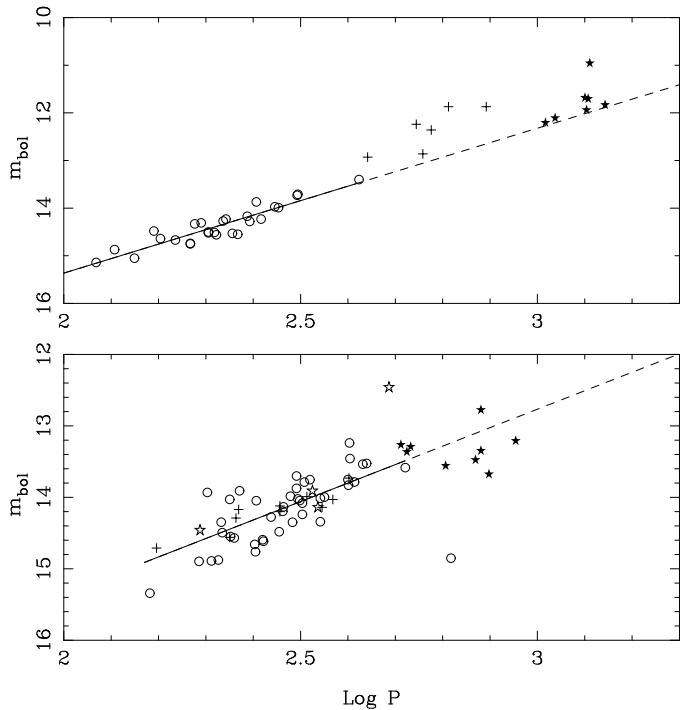,width=9.0cm,angle=-90}}
\vspace{-0.3cm}
\caption[]{Period-Luminosity relation for O-rich (upper panel) and
C-rich (lower panel) AGB stars in the LMC. The solid and dashed lines
are least-square fits (and the extrapolations, respectively) to the data
points represented by the open symbols in the upper panel and open
symbols and crosses in the lower one (except the two obvious
outliers). The filled symbols represent IRAS detected stars. Taken
from Groenewegen et al. (1996), where more details are provided.
}
\vspace{-0.3cm}
\end{figure*}

\vspace{-0.3cm}
\section{First DENIS results}

DENIS (Deep Near-Infrared Survey) is a survey of the southern sky in
$IJK$, to limiting 3$\sigma$ magnitudes of 18.5, 16.3, 14.0. It uses
the 1m telescope at La Silla, Chile. More information can be found in
Epchtein (1997), Epchtein et al. (1997), and in Skrutskie et al. (1997) on a
similar US all-sky $JHK$ survey called 2MASS. Figure~3 shows some
DENIS results of two fields in the LMC.

\begin{figure*}
\centerline{\psfig{figure=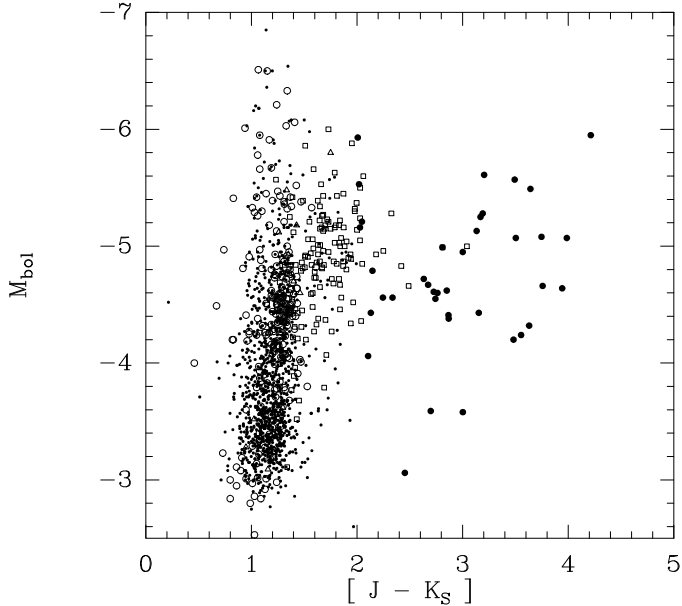,width=9.0cm
}}
\caption[]{DENIS observations of two 0.18 sq. deg. fields covering
part of the Blanco et al. (1980) fields. $M_{\rm bol}$ is based 
on $K$ and $J-K_s$. Plotted are the 1200 objects with errors in their
colors smaller than 0.15 mag. Open circles represent known M-stars,
open squares known C-stars, open triangles known LPVs of unknown
spectral type. Solid points are new DENIS sources.
From Loup et al. (1997b), where more details are provided.
}
\vspace{-0.3cm}
\end{figure*}

\vspace{-0.3cm}
\section{First ISO results}

Almost all ISO observations of AGB stars in the MCs are being done
with PHT and CAM, as the objects are to faint to be observable with
SWS and LWS. There is one exception and that is the supergiant WOH G064
that will be observed with SWS.  For other observations of this
interesting object see the poster paper by van Loon in this volume.

\noindent
Guaranteed and open time proposals regarding the MCs are as follows.
\vspace{-0.3cm}

\begin{itemize}

\item PHT

There are the proposals by Trams et al. (see his contribution in this
volume) and Blommaert et al., who
focus on individual IRAS detected stars in the LMC and SMC,
respectively.  Some PHT-S spectra are obtained for the LMC sources,
and both projects aim to observe at 12, 25 and 60 $\mu$m (see Fig.~4
for an example of an ISO 60 $\mu$m observation).

\item CAM

There are complementary programs by the same two authors to complete
the SEDs at shorter wavelengths. In addition CVF spectra are obtained
for some sources (see Blommaert's contribution in this volume).

There are two mini-surveys being carried out: by Tanab\'e et al. using
3 filter observations of 18 globular clusters (see the contribution in
this volume), and by Loup et al. using 2 filter observations of 7
fields in the LMC for a total 0.5 sq. deg. (see the poster paper by
Josselin et al. in this volume).

\end{itemize}
\vspace{-0.3cm}

\noindent
The most exciting to come out of these ISO results so far is the
detection of extremely red, relatively low luminosity objects. The
most extreme example is shown in Fig.~5. This SMC star was not
detected by us in
$K$ down to 14$th$ magnitude. Fitting its SED results in a luminosity of 5300
\lsol\ and a estimated mass loss rate of 7 $\times$ 10$^{-5}$
\msolyr. Its optical depth is as large as that of the reddest known Galactic
carbon Mira, which happens to have also the longest known period
for a carbon Mira. However, that object, using the $P-L$ relation has
a luminosity of 16~400 \lsol\ (!), whereas the mass loss rate is comparable.

\begin{figure*}
\centerline{\psfig{figure=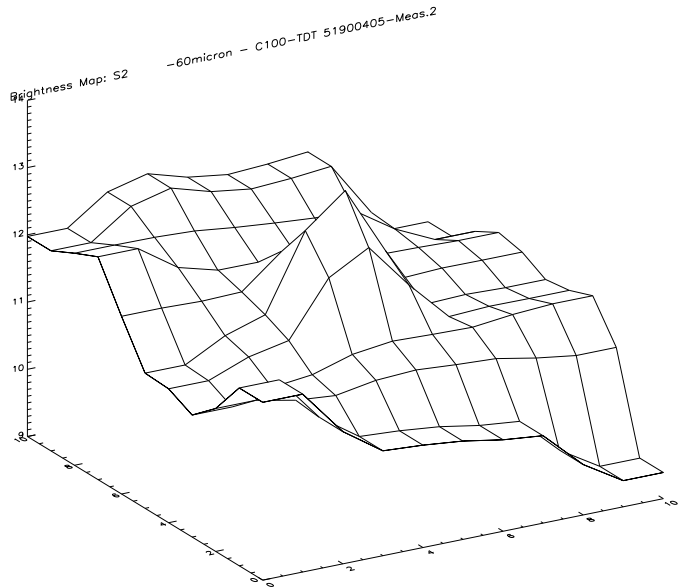,width=10.0cm
}}
\vspace{-0.3cm}
\caption[]{ISOCAM 3$\times$3 raster map observations with C100 at 
60 $\mu$m of a star in the SMC. The preliminary analysis results in 
a flux of 59 mJy. The axes represent surface brightness versus pixels.
}
\vspace{-0.4cm}
\end{figure*}

\begin{figure*}
\centerline{\psfig{figure=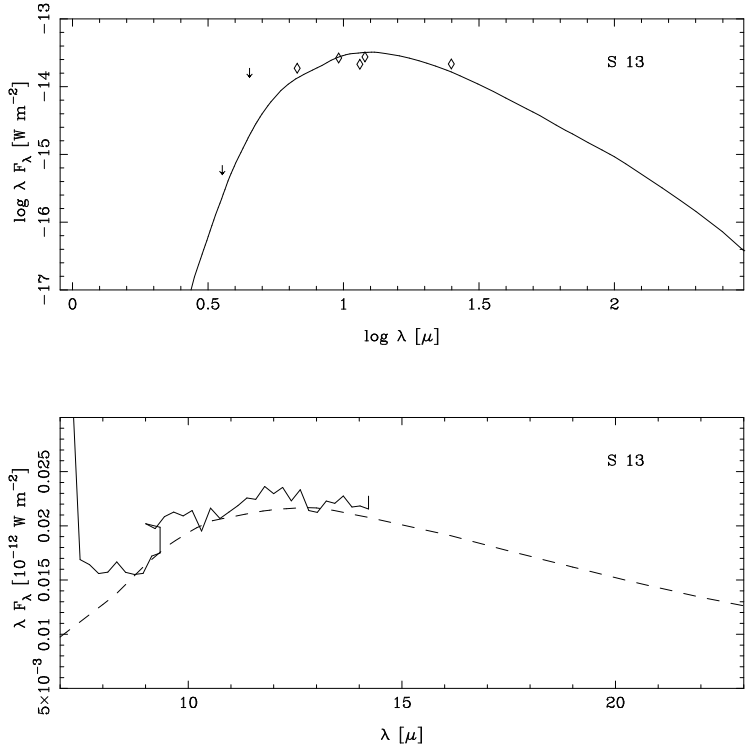,width=10.0cm,angle=-90}}
\caption[]{Model fit to the ISOCAM broad-band and CVF spectra of the
object S13.  It is as red as the reddest known carbon star in our
Galaxy. Amorphous carbon dust is assumed. The sharp drop of flux below
8 $\mu$m in the CVF spectrum is instrumental.
}
\vspace{-0.4cm}
\end{figure*}

\vspace{-0.3cm}
\section{Conclusions}

ISO provides the opportunity to observe AGB stars in the MCs with
unprecedented sensitivity and at wavelengths that are inaccessible
from the ground. It is in a way unfortunate that the majority of ISO
time is spent on follow-up observations of IRAS detected stars, as we
know that this sample is biased. Only one program uses the survey
capability of CAM to observe field stars, but even that is restricted
to 0.5 sq. degree in the LMC only.

There is life after ISO: the near-infrared surveys DENIS and 2MASS
will observe the whole of the MCs, and in this respect will provide
the most unbiased view of the AGB population.
In addition, the  multi-wavelength photometric and spectroscopic
capabilities of the VLT, the multi-object optical spectroscopic
capability of de 2dF instrument on the AAT, and the MACHO/EROS
projects that identify LPVs ensure that extra-galactic AGB research
will be an interesting research topic in the future.

\vspace{-0.3cm}
\subsection*{Acknowledgements}

I would like to thank the people who have feeded me with data and for
useful discussions and without whom this review would not have been
possible: Joris Blommaert, Maria Rosa Cioni, Eric Josselin, Jacco van
Loon, Cecile Loup and Norman Trams.

\end{document}